# Energetics and kinetics of Li intercalation in irradiated graphene scaffolds

J. Song[1, a)], B. Ouyang[1], N. V. Medhekar[2, b)]

1. Department of Mining and Materials Engineering, McGill University, Montreal, QC, Canada

2. Department of Materials Engineering, Monash University, Clayton, Victoria, Australia

## Abstract

In the present study we investigate the irradiation-defects hybridized graphene scaffold as one potential building material for the anode of Li-ion batteries. Designating the Wigner $V_2^2$ defect as a representative, we illustrate the interplay of Li atoms with the irradiation-defects in graphene scaffolds. We examine the adsorption energetics and diffusion kinetics of Li in the vicinity of a Wigner $V_2^2$ defect using density functional theory calculations. The equilibrium Li adsorption sites at the defect are identified and shown to be energetically preferable to the adsorption sites on pristine (bilayer) graphene. Meanwhile the minimum energy paths and corresponding energy barriers for Li migration at the defect are determined and computed. We find that while the defect is shown to exhibit certain trapping effects on Li motions on the graphene surface, it appears to facilitate the interlayer Li diffusion and enhance the charge capacity within its vicinity because of the reduced interlayer spacing and characteristic symmetry associated with the defect. Our results provide critical assessment for the application of irradiated graphene scaffolds in Li-ion batteries.

---

[a)] Author to whom correspondence should be addressed. E-Mail: jun.song2@mcgill.ca
[b)] Email: nikhil.medhekar@monash.edu

**Introduction**

Carbon based materials such as graphite and multilayer graphene have drawn paramount of interests as the state-of-art anode material for Li ion batteries [1, 2, 3, 4, 5]. These materials exhibit good Li storage capacity and fast charge/discharge characteristics [6, 7, 8, 9, 10]. They also possess high electronic/thermal conductivity and fast Li diffusion that are especially advantageous for battery applications [11, 12, 13]. In addition, they exhibit excellent in-plane mechanical integrity and are often combined with other high-capacity anode materials (e.g., Si, $FePO_4$, and *etc.*) [14, 15, 16] to improve the failure resistance and cycling performance of Li-ion batteries.

One major challenge associated with intercalation based Li-ion batteries is the repeated volume expansion/shrinkage during Li insertion/extraction. The volume change can be substantial (*i.e.* >150%) for high-capacity Li-ion batteries [17, 18], leading to substantial stress that causes electrode fracture, loss of electrical contact and rapid fading of capacity. Graphite/multilayer graphene, despite their in-plane strength, are also vulnerable to large volumetric deformations because of the weak bonding (van der Waals forces) between graphene sheets, and often suffer interlayer failure during Li intercalation. For instance, one major cause of degradation in graphite-based Li-ion batteries is the constant breakage and reformation of the solid electrolyte interfaces [19, 20] due to exfoliation [21, 22].

The above challenges pertinent to graphite/multilayer graphene may however be potentially overcome by fabricating hybrid 3D graphene scaffolding structures where graphene sheets are "cross-linked" through covalent bonds between neighboring sheets, as illustrated in Figure 1a. The idea of such scaffolding structures has been previously demonstrated by da Silva *et al.* [23] for carbon nanotube bundles (CNBs) where covalent bonds construct strong cross-links between individual nanotubes, leading to a sizable increase in the shear modulus of CNBs. It has been shown [23, 24, 25, 26, 27, 28] that the covalent cross-links in carbon based materials can be introduced via high energy defects produced by irradiation [29, 30, 31]. In particular for graphite/multilayer graphene, one prevailing category of such defects are the Wigner defects [32]. The Wigner defects exhibit formation energies and migration barriers on the order of ~10 eV, making them thermodynamically stable and immobile at ambient temperature. However the presence of those defects necessarily modifies the local atomic and electronic structures, and subsequently how Li atoms interact with graphite/multilayer graphene. In this regard, it is of great importance to understand the interactions between Li atoms and the Wigner defects, and how their interactions impact the Li transport process and consequently battery performance.

In the present study, we focus on a AB-stacked bilayer graphene with a Wigner $V_2^2$ defect [23] as a simple version of the graphene scaffold to illustrate the role of irradiation-defects during Li intercalation. Through density functional theory (DFT) calculations, the equilibrium adsorption sites and energetics of Li atoms, in the low concentration limit, in the vicinity of the defect are identified. Meanwhile the charge transfer processes and electronic structures for various Li-defect configurations are investigated. Then the local migration paths/barriers of Li at the defect are examined. In the end the implications of our findings on Li transport in irradiated graphene scaffolds and subsequently the charging behaviors of Li-ion batteries are discussed.

1. **Computational methodology**

Spin polarized DFT calculations were performed using the Vienna ab-initio Simulation Package (VASP)[33] with projector augmented-wave (PAW) potentials [34, 35]. The electrons explicitly included in the calculations are the ($2s^2 2p^2$) electrons of carbon and the ($1s^2 2s^1$) electrons of Li. A cutoff energy of the plane wave basis set of 800 eV is used in all calculations. The Wigner $V_2^2$ defect examined in this study is a crossplanar divacancy formed as a result of the coalescence of two interplanar vacancies. It exhibits lower formation energy and thus thermodynamically more stable than other Wigner divacancy defects. Nonetheless preliminary studies on the other Wigner divacancy defects (e.g., Wigner $V_2^1$) have also been performed, showing largely similar results (*i.e.*, in terms of Li energetics and kinetics at the defect).

In all calculations, a simulation cell with a Wigner $V_2^2$ defect centered in a $4 \times 4 \times 1$ bilayer graphene is created, as shown in Figure 1. Sample calculations using larger cell sizes (i.e., up to $8 \times 8 \times 1$ cell) are also performed, showing no size dependence of our results. The distance between two neighboring bilayer graphenes is chosen to be 15Å to eliminate image interactions across the periodic boundary perpendicular to the graphene sheet. The simulation cell is first relaxed with both the cell shape and volume allowed to change to reach the ground state, following which Li atoms are introduced individually to examine their adsorption at the defect. In the relaxations of ionic coordinates and supercell vectors, the convergence was considered reached when the forces on all ions were less than 0.01 eV/Å. Then the migration kinetics of Li between neighboring adsorption sites are investigated using the climbing image Nudged Elastic Band (ci-NEB) method [36]. The NEB calculation is considered converged when the force on each image is less than 0.01 eV/Å.

One thing to note is that the van der Waals (vdW) interactions were not considered in the present study. Previously it was shown by Lee et al.[37] and Fan et al.[38] that the vdW interactions can have sizeable effects on the Li adsorption energy on graphene. However we find that the vdW interactions have rather small influence on Li adsorption around the Wigner defect (see Supplementary Information). This is likely due to the interlayer bond at the defect. The interlayer bond, being covalent in nature, dominates the local atomic structure at the defect and thus the Li energetics.

## 2. Results and discussion

### 2.1. Benchmark results of the Wigner $V_2^2$ defect

To put the interplay between the Wigner $V_2^2$ defect and Li atoms into proper perspective, we first examine the structural and electronic properties of the Wigner defect in absence of Li. The relaxed atomic configuration of the defect is shown in Figure 1b. The graphene sheets at the defect are interconnected via a covalent bond of 1.38 Å in length. C atoms reconstruct at the vicinity of the defect, exhibiting considerable out-of-plane displacements. These results are in excellent agreement with those previously reported in Ref. [32].

The significant geometric reconstruction induced by the defect necessarily modifies the local electronic properties. In this regard, we examine the band structure and density of states (DOS) of the system, shown in Fig. 1c. We note from Fig. 1c that the system exhibits a Fermi energy $E_f$ = -0.438 eV that is about 0.08 eV lower than the pristine bilayer graphene. Also we see that there is a noticeable band gap of $\Delta E \sim 0.08$ eV [39] at the Fermi level in comparison to the near-zero band gap in the pristine bilayer graphene. Analysis of the DOS shows that the bands near the Fermi level come from those reconstructed C atoms in the immediate vicinity of the defect, suggesting that the band gap (or doping level in the case of dilute defect concentration, see note in Ref. [39]) directly attributes to the Wigner defect.

## *2.2. Energetics and electronic structures of Li atoms at a Wigner $V_2^2$ defect*

Li atoms, when intercalated into a AB-stacked bilayer graphene (with or w/o the Wigner $V_2^2$ defect), may either occupy sites on the outer surfaces or between two graphene layers, denoted as T-sites and M-sites respectively. We note that the locations of these sites can be different as the Li concentration varies[40]. For simplicity in the context below, we limit our discussion to the low Li concentration regime.

For a pristine AB-stacked bilayer graphene, the Li adsorption on the outer surfaces is similar to the case of a monolayer graphene, with the equilibrium T-sites being directly on top of the hexagon (hollow sites) instead of on top of C atoms (i.e., top sites) or on top of C-C bonds; while the equilibrium M-sites are hollow sites *w.r.t* one graphene sheet and top sites *w.r.t* the other graphene sheet [28, 41, 42]. To assess the energetics of Li adsorption on bilayer graphene, we compute the adsorption energy $E_{ad}$ of a Li adatom as:

$$E_{ad} = E_{\text{Li-GP}} - E_{\text{Li}} - E_{\text{GP}},  \qquad (1)$$

where $E_{\text{Li-GP}}$, $E_{\text{Li}}$ and $E_{\text{GP}}$ denote the total energy of Li adsorbed bilayer graphene, the energies of an isolated Li atom and the bilayer graphene respectively. For an AB-stacked bilayer graphene, we find that the adsorption energies of Li at T-sites and M-sites are -1.36eV and -2.15eV with the Li adatom being 1.73Å and 1.92Å away from the corresponding hexagon centers respectively, as listed in Table I.

Locally at the defect, the interactions between Li and C atoms are however necessarily different from the case of a pristine bilayer graphene. Figure 2 shows the positions of a subset of the equilibrium Li adsorption sites identified in the vicinity of the Wigner $V_2^2$ defect, showing six non-identical T-sites and five M-sites that correspond to local energy minimums. These T-sites and M-sites are numbered from 1-6 and 1-5 respectively. We see from Figure 2 that among the T-sites, site 1 is centered directly above the defect, sites 2-5 sit roughly on top of hexagons and site 6 sits above a pentagon; while for the M-sites, site 1 sits above a pentagon and the rest stay on top of hexagons with respect to the bottom sheet. These sites also present the "*basis*" set, from which other equivalent adsorption sites around the defect can be identified via symmetry operations. The corresponding adsorption energies of Li at those sites are listed in Table I. We note that for both the T-sites and M-sites, the $E_{ad}$s of Li at the defect are lower than the ones in a pristine AB stacked bilayer graphene, meaning Li is more energetically favorable around the defect. Thus there is a thermodynamic tendency for intercalated Li atoms to segregate at Wigner defects. To understand the strong binding of Li atoms at the Wigner defect, below we examine the local electronic structure and charge transfer phenomena.

Figure 3 shows two representative atomic configurations, with Li adsorbed at T-site 1 (cf. Fig. 3a) and M-site 1 (cf. Fig. 3b), along with the corresponding charge distributions that are represented by the charge density difference $\Delta\rho$, defined as:

$$\Delta\rho = \rho_{Li+GP} - [\rho_{Li} + \rho_{GP}], \qquad (2)$$

where $\rho_{Li+GP}$, $\rho_{Li}$ and $\rho_{GP}$ denote the charge densities for the whole system, an isolated Li atom and the defective bilayer graphene respectively. The $\Delta\rho$ contours are drawn on top of the atomic configurations in Figure 3. In both cases we see significant charge transfer from the Li atom to its neighboring C atoms. In order to quantify the amount of charge transfer between Li and C atoms, we perform the Barder charge analysis [43, 44, 45] which computes the estimate of charge on each atom. For the particular configuration shown in Fig. 3a where Li occupy T-site 1, the Barder analysis shows a total charge of 2.1 electrons on Li, suggesting Li very much completely lose its valence electron. The charge Li loses is then found to distribute mostly among its immediate C neighbors, as shown in Fig. 3c where the amount of charge transferred to each C atom is indicated. Similar analysis has been performed for the case of Li occupying M-site 1. We again found that the Li atom gives away its valence electron, distributed among its immediate C neighbors. However in case of Li being at M-site 1, charge gets transferred to C atoms in both graphene sheets. We note that in both cases shown in Figs 3a-b, the distribution of charge is largely non-uniform among C atoms neighboring Li, in sharp contrast to the case of Li sitting on a pristine (bilayer) graphene where the transferred charge is mostly distributed uniformly among the six C atoms immediately surrounding Li [46].

Meanwhile the corresponding band structures and DOSs of Li-defect systems are computed. Figure 4 shows the results for the two sample configurations previously shown in Figs 3a-b. We see that the Fermi energies are -0.0872 eV (at the T-site) and -3.072 eV (at the M-site) respectively, both higher than the one of a Li-free Wigner $V_2^2$ defect. In addition we observe that several bands above the Fermi level shift downwards upon Li adsorption, accompanying which the band gap disappears, suggesting that the material changes from a semiconductor to a conductor. Similar trends are also observed for Li adsorbed at other locations. These findings resonate with the case of Li adsorption on pristine (bilayer) graphene where zero band gaps and increase in the Fermi energies are also observed. However we note that the change in the Fermi level induced by Li adsorption at the defect is much smaller [41, 47, 48].

## 2.3. *Migration properties of Li at a Wigner $V_2^2$ defect*

Above we have demonstrated that Li atoms exhibit lower adsorption energies (cf. Table I) at the Wigner defect. This is likely due to the defect modifying the local charge distribution/electronic structure, and subsequently Li-C interactions[41, 49]. The preferential binding suggests segregation and potential trapping of Li at the defect. This necessarily impacts the local Li kinetics and subsequently battery charging behaviors. For example, the Wigner defect may serve as a potential *sink* to trap Li motions, which can result in advocate effects in battery capacity and charging/discharging rates. In this regard we investigate the migration properties of Li at the Wigner defect. From Table I we note that Li has the lowest adsorption energies at the T-site 1 and M-site 1 (i.e., when adsorbed on outer graphene surface and in the middle respectively). Thus they represent the strongest *trapping* sites at the defect. In the follows we examine the migration of Li from those sites to other equilibrium binding sites in their immediate neighborhoods. The minimum energy paths (MEPs) are examined through NEB calculations with T-site 1 or M-site 1 taken as the initial point and its immediate neighboring binding sites taken as final points. Depending on the distance between the initial and final points, 5-10 images are used in the NEB calculation. The corresponding migration barriers (denoted as $E_m$) obtained are listed in Table II, and two representative MEPs are shown in Figure 5.

For Li kinetics starting from T-site 1, the final point can be T-sites 2-6. Among the MEPs we find that the saddle point occurs on top of a C atom for Li migrating towards T-sites 2-3, and at a bridge site above the C-C bond for Li migrating towards T-sites 4-6. The corresponding $E_m$ ranges from 0.26 to 0.54eV (cf. Table II). We note that these $E_m$ values are on the same order as the $E_m$ (i.e., spanning from 0.322 to 0.34eV) of Li on the outer surface of a graphene or graphite[50, 51] though on average higher. Also we note that the MEP for Li migration from T-site 1→T-site 2 yields an $E_m$ of 0.26eV, suggesting that Li can easily escape from the Wigner defect through that path.

On the other hand, for interlayer Li kinetics at the Wigner defect the migration does not occur directly between the M-sites identified. Instead a MEP starting from a M-site would end at an *image* M-site. Those *image* M-sites are a set of Li adsorption sites that are equivalent images of the M-sites listed in Table I due to the Ci (i.e., inverse) symmetry of the Wigner $V_2^2$ defect [52]. For instance, the MEP starts from the M-site 1 would end at a site (denoted as site I5 in Figure 5) that is an equivalent image of the M-site 5, as illustrated in Figs 5b and 5d. Overall our findings indicate that the $E_m$ between M-sites (and *image* M-sites) at the Wigner defect is in the order of 0.08 eV (e.g., $E_m$ is 0.083 eV for migration from M-site 1 to image M-site I5, cf. Fig. 5b and 5d), significantly lower than the migration barrier for Li motions on the outer graphene surface. This value of $E_m$ is also lower than the interlayer Li migration barrier in graphite (~0.2eV) [53]. The low $E_m$ presumably comes from the reduced interlayer separation and AB-stacking locally at the defect. In addition, the presence of the *image* M-sites effectively doubles the Li adsorption sites and thus the charge capacity compared to the normal Li-graphite system. We note that however Li intercalation will induce the overall sequence of graphene sheets to transit from AB-stacking to AA-stacking [7]. Thus a competition between the stacking sequences is expected during lithiation. Nonetheless the atomic configuration in the immediate vicinity of the Wigner defect likely can be (partially) retained because of the strong interlayer covalent bonding.

From our results on Li migration, we note that while the Wigner $V_2^2$ defect has some adverse effects on Li migration on the graphene surfaces (cf. Table II), it facilitates Li diffusion and increases local charge capacity for interlayer Li intercalation because of the defect-induced stacking sequence and reduction in interlayer spacing. In addition it was postulated by several studies [37, 54, 55] that structural defects, particularly vacancy-like defects, can aid Li diffusion through graphene sheets. Though the through-layer Li kinetics is beyond the scope of our current study, we have performed some preliminary calculations that show the diffusion barrier for through-layer Li diffusion via the Wigner defect is 7-8 eV, being 2-3 eV smaller than the diffusion through a pristine graphene sheet[37]. Thus the effects of the Wigner defect are two-fold. Given that the overall Li kinetics is a combination of contributions from surface, interlayer and through-layer Li diffusion, one may tune the defect density together with the surface-to-volume ratio to optimize the performance of 3D graphene scaffolds in Li-ion batteries.

## 3. Conclusions

To conclude, we have performed first-principles calculations to study the adsorption energetics and transport kinetics of Li in the vicinity of a Wigner $V_2^2$ defect to illustrate the interplay between Li atoms and irradiation-induced defects in graphite/multilayer graphene. We have identified equilibrium adsorption sites for Li at the Wigner defect and shown that they are energetically preferable compared to the sites on pristine (bilayer) graphene. We find that the migration barriers for Li diffusion on the graphene surface increase at the vicinity of the defect, suggesting adverse effects of defects on surface Li motions. On the other hand the interlayer Li diffusion along with the charge capacity are enhanced around the defect, because of the reduced interlayer spacing and particular symmetry associated with the defect. Our study provides important insights toward the design and application of irradiation-defects hybridized graphene scaffolds in Li-ion batteries.

**Acknowledgement:** The authors acknowledge support of this work by the NSERC Discovery grant (grant # RGPIN 418469-2012). The authors also would like to acknowledge Supercomputer Consortium Laval UQAM McGill and Eastern Quebec as well as MASSIVE and NCI computing infrastructure provided by the Australian Government for providing computing power.

**Table I**: The computed adsorption energies at different sites.

| T-Sites    | 1     | 2     | 3     | 4     | 5     | 6     | $H_t^*$ |
|------------|-------|-------|-------|-------|-------|-------|---------|
| $E_{ad}$ (eV) | -2.26 | -2.00 | -1.86 | -1.86 | -1.98 | -2.09 | -1.36   |
| M-Sites    | 1     | 2     | 3     | 4     | 5     | $H_m^*$ |         |
| $E_{ad}$ (eV) | -2.60 | -2.53 | -2.47 | -2.34 | -2.54 | -2.15 |         |

* $H_t$ and $H_m$ denote the hollow sites on out surface or between two graphene sheets in a pristine AB stacked bilayer graphene respectively.

Table II: The migration barrier $E_m$ of Li

| Path       | T-site 1→2 | T-site 1→3 | T-site 1→4 | T-site 1→5 | T-site 1→6 |
|------------|------------|------------|------------|------------|------------|
| $E_m$ (eV) | 0.260      | 0.480      | 0.540      | 0.498      | 0.436      |
| Path       | Adjacent sites on surface of AB-stacked bilayer graphene ||||| 
| $E_m$ (eV) | 0.33$^a$, 0.322$^b$ and 0.34$^c$ |||||
| Path       | M-site 1→I5 |||||
| $E_m$ (eV) | 0.083       |||||

*a*: This work; *b*: Refs [50]; *c*: Ref [46].

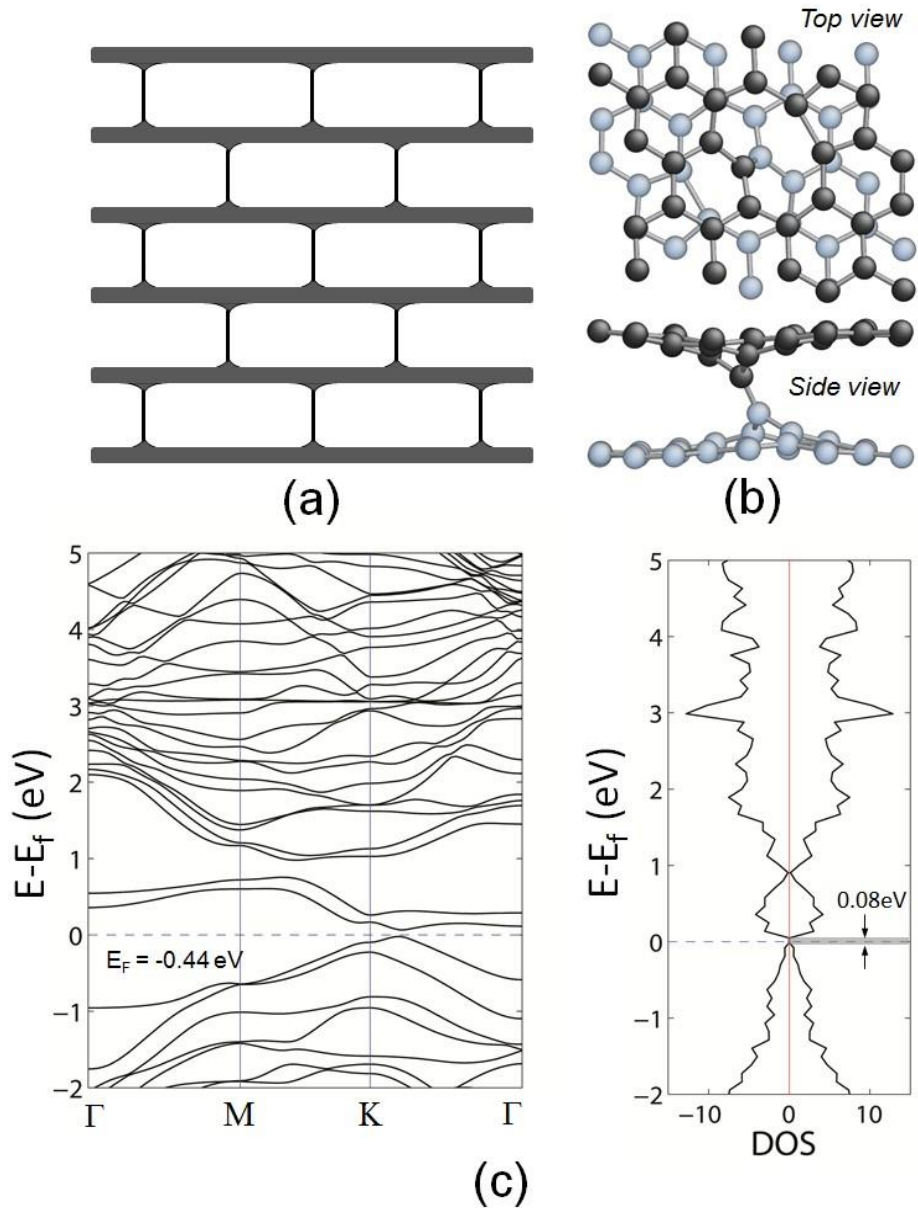

**Figure 1 (color online)**: *a*) The illustration of an example 3D graphene scaffold structure where neighboring graphene sheets (i.e., colored dark gray) are bridged through covalent links (i.e., indicated by vertical line segments) provided by crossplanar defects; *b*) top and side projection views of the Wigner $V_2^2$ defect with the C atoms in different graphene sheets colored dark gray and cyan for clarity. The corresponding band structure and density of states (DOS) plots of the system are shown in *c*) where a band gap opening of 0.08 eV is indicated.

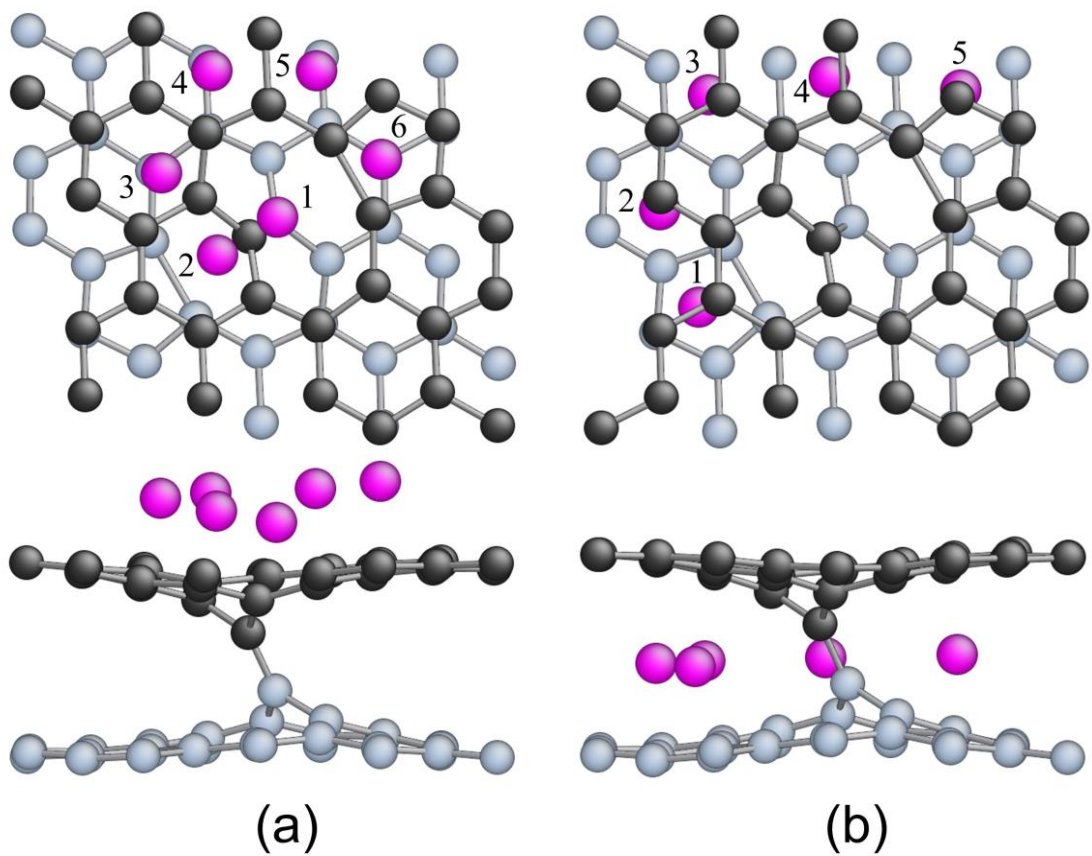

**Figure 2 (color online)**: The top and side views of the nonequivalent (*a*) T-sites 1-6 on graphene surfaces, and (*b*) M-sites 1-5 between graphene sheets for Li adsorption. Here the adsorption sites are colored magenta, while the C atoms in different graphene sheets are colored dark gray and cyan respectively for clarity.

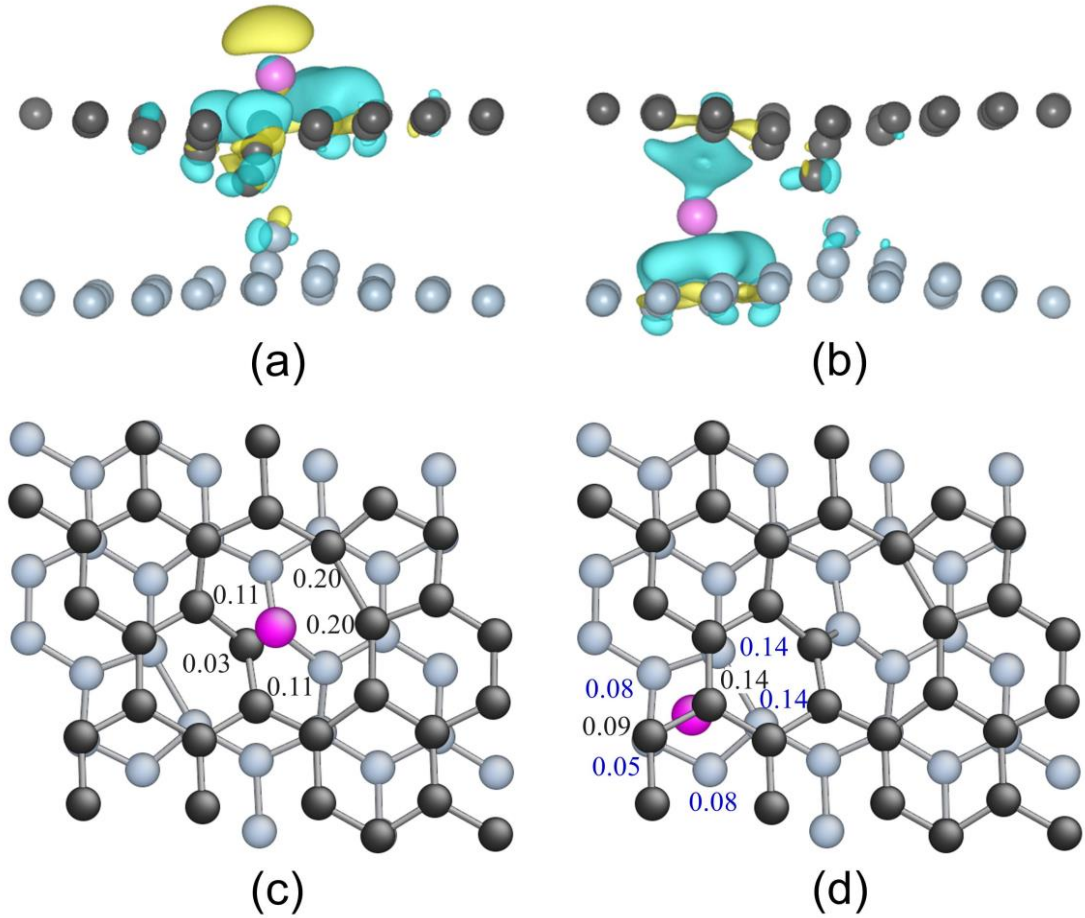

**Figure 3 (color online)**: The side views of the charge difference (cf. Eq. 2) contours around the Wigner $V_2^2$ defect when a Li atom is adsorbed (*a*) at T-site 1 and (*b*) M-site 1, with the Li atom colored magenta while C atoms in the top and bottom graphene sheets colored dark gray and cyan respectively. The amounts of charge transfer in the unit of electron from the Li atom to the C atoms neighboring it are indicated in the corresponding top projection views (*c*) and (*d*). The black and blue numbers indicate charge transferred to C atoms on the top and bottom sheets respectively.

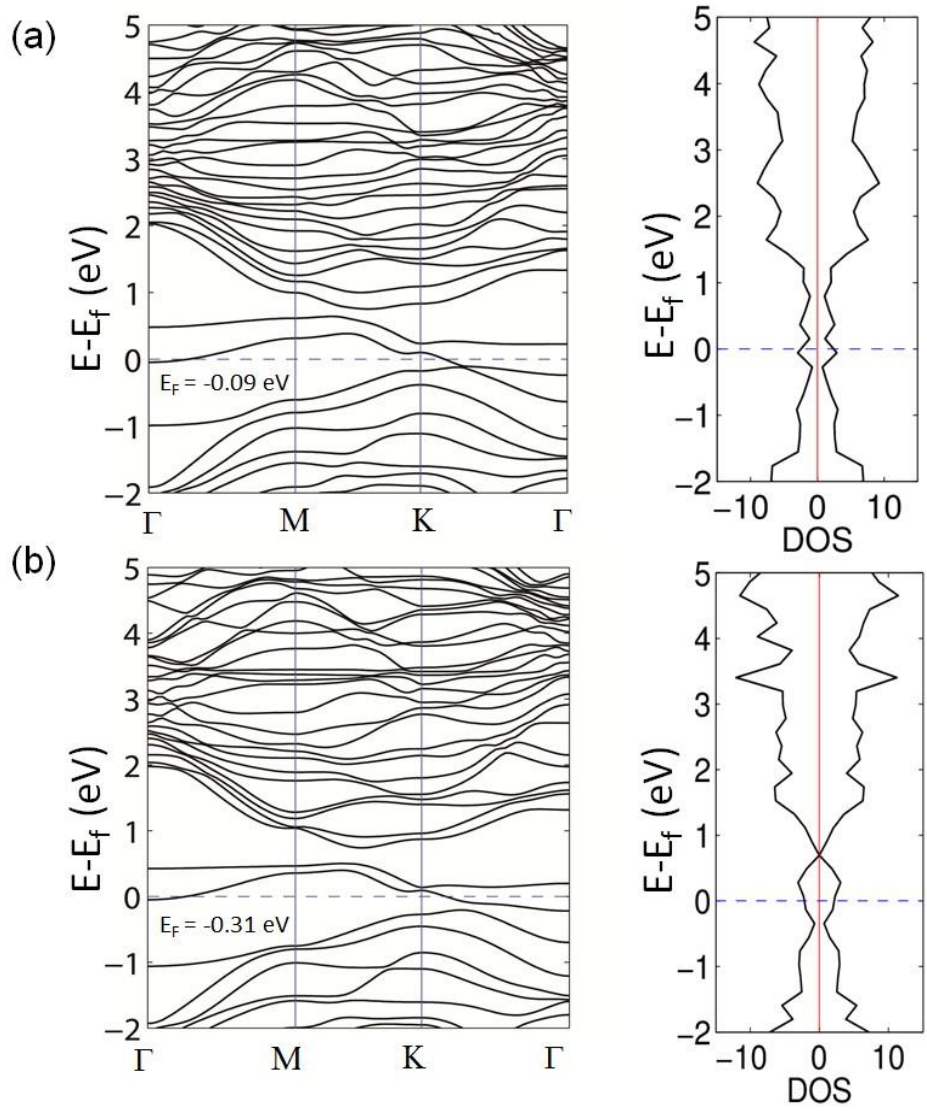

**Figure 4**: The plots of band structure and density of states (DOS) of the Li-defect systems where a Li atom is adsorbed at (*a*) T-site 1 and (*b*) M-site 1 respectively.

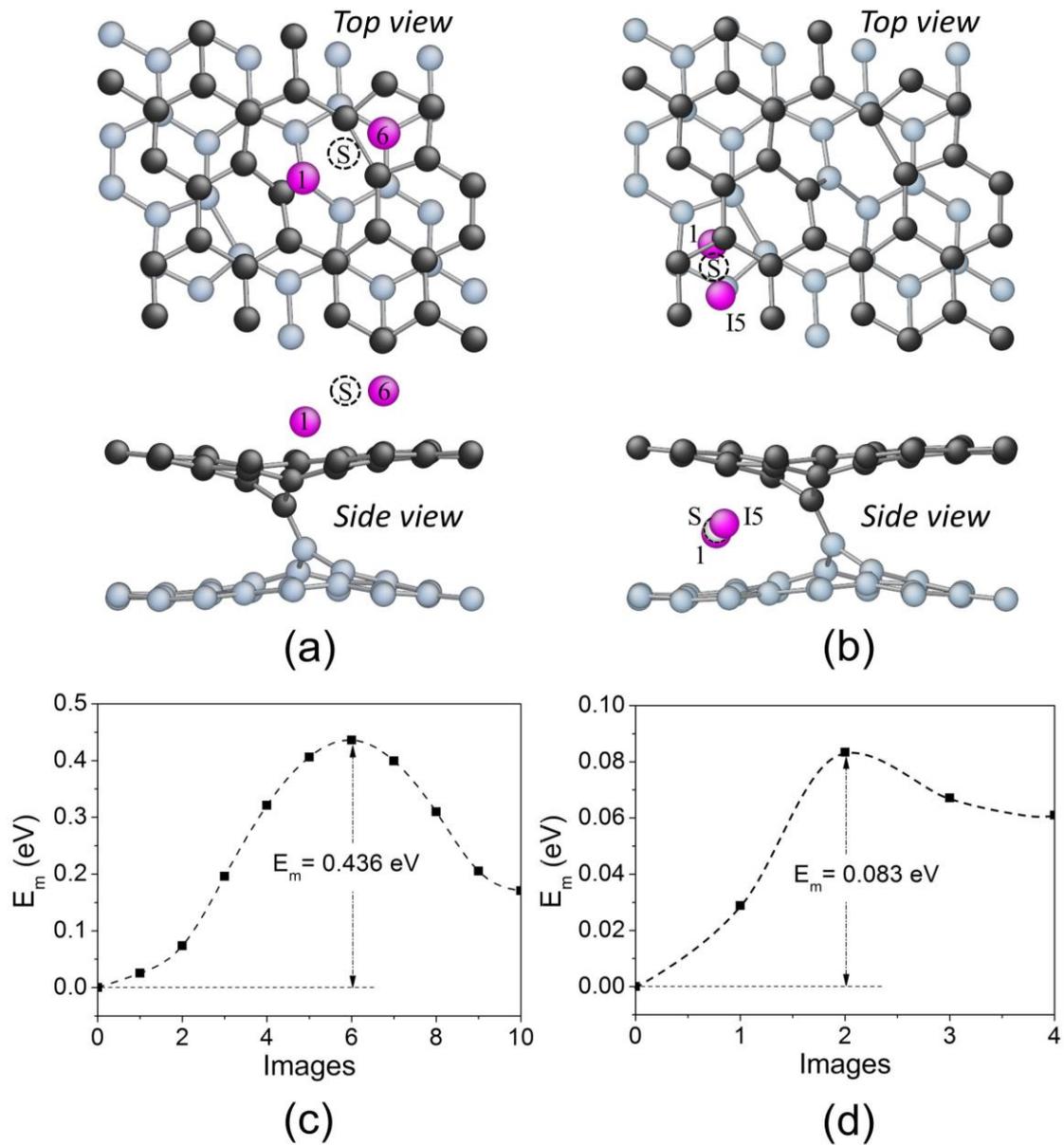

**Figure 5 (color online)**: The migration paths of Li (identified via the NEB calculations) (*a*) from T-site 1 to T-site 6 and (*b*) from M-site 1 to its neighboring image site I5, with the Li positions corresponding to the saddle points indicated by the dashed white circles. The corresponding MEPs are shown in (*c*) and (*d*) respectively.

**References**

(1) Dresselhaus, M. S.; Dresselhaus, G. *Adv. Phys.* **1981,** *30,* 139-326.
(2) Pistolia, G., *Lithium Batteries: New Materials, Developments, and Perspectives*. Elsevier: New York, 1994.
(3) Dahn, J. R.; Zheng, T.; Liu, Y. H.; Xue, J. S. *Science* **1995,** *270,* 590-593.
(4) Fischer, J. E. *Chemical Innovation* **2000,** *30,* 21-27.
(5) Juza, R.; Wehle, V. *Naturwissenschaften* **1965,** *52,* 560.
(6) Kaskhedikar, N. A.; Maier, J. *Adv. Mater.* **2009,** *21,* 2664-2680.
(7) Winter, M.; Besenhard, J. O.; Spahr, M. E.; Novak, P. *Adv. Mater.* **1998,** *10,* 725-763.
(8) Kasuh, T.; Mabuchi, A.; Tokumitsu, K.; Fujimoto, H. *J. Power Sources* **1997,** *68,* 99-101.
(9) Azuma, H.; Imoto, H.; Yamada, S.; Sekai, K. *J. Power Sources* **1999,** *82,* 1-7.
(10) Tokumitsu, K.; Fujimoto, H.; Mabuchi, A.; Kasuh, T. *Carbon* **1999,** *37,* 1599-1605.
(11) Castro Neto, A. H.; Guinea, F.; Peres, N. M. R.; Novoselov, K. S.; Geim, A. K. *Rev. Mod. Phys.* **2009,** *81,* 109-162.
(12) Geim, A. K.; Novoselov, K. S. *Nat. Mater.* **2007,** *6,* 183-191.
(13) Balandin, A. A.; Ghosh, S.; Bao, W. Z.; Calizo, I.; Teweldebrhan, D.; Miao, F.; Lau, C. N. *Nano Lett.* **2008,** *8,* 902-907.
(14) Evanoff, K.; Magasinski, A.; Yang, J. B.; Yushin, G. *Adv. Energy Mater.* **2011,** *1,* 495-498.
(15) Magasinski, A.; Dixon, P.; Hertzberg, B.; Kvit, A.; Ayala, J.; Yushin, G. *Nat. Mater.* **2010,** *9,* 353-358.
(16) Lung-Hao Hu, B.; Wu, F.-Y.; Lin, C.-T.; Khlobystov, A. N.; Li, L.-J. *Nat. Commun.* **2013,** *4,* 1687-1693.
(17) Boukamp, B. A.; Lesh, G. C.; Huggins, R. A. *J. Electrochem. Soc.* **1981,** *128,* 725-729.
(18) Huggins, R. A. *J. Power Sources* **1999,** *81,* 13-19.
(19) Balbuena, P. B.; Wang, Y., *Lithium-Ion Batteries: Solid-Electrolyte Interphase*. Imperial College Press: London, 2004.
(20) Peled, E. *J. Electrochem. Soc.* **1979,** *126,* 2047-2051.
(21) Aurbach, D.; Markovsky, B.; Weissman, I.; Levi, E.; Ein-Eli, Y. *Electrochim. Acta* **1999,** *45,* 67-86.
(22) Gnanaraj, J. S.; Levi, M. D.; Levi, E.; Salitra, G.; Aurbach, D.; Fischer, J. E.; Claye, A. *J. Electrochem. Soc.* **2001,** *148,* A525-A536.
(23) da Silva, A. J. R.; Fazzio, A.; Antonelli, A. *Nano Lett.* **2005,** *5,* 1045-1049.
(24) Banhart, F. *Rep. Prog. Phys.* **1999,** *62,* 1181-1221.
(25) Kis, A.; Csanyi, G.; Salvetat, J. P.; Lee, T. N.; Couteau, E.; Kulik, A. J.; Benoit, W.; Brugger, J.; Forro, L. *Nat. Mater.* **2004,** *3,* 153-157.
(26) Peng, B.; Locascio, M.; Zapol, P.; Li, S. Y.; Mielke, S. L.; Schatz, G. C.; Espinosa, H. D. *Nat. Nanotechnol.* **2008,** *3,* 626-631.
(27) Sammalkorpi, M.; Krasheninnikov, A.; Kuronen, A.; Nordlund, K.; Kaski, K. *Phys. Rev. B* **2004,** *70,* 245416.

(28) Huhtala, M.; Krasheninnikov, A. V.; Aittoniemi, J.; Stuart, S. J.; Nordlund, K.; Kaski, K. *Phys. Rev. B* **2004,** *70,* 045404.

(29) Iwata, T. *J. Nucl. Mater.* **1985,** *133,* 361-364.

(30) Kelly, B. T., *The physics of Graphite*. Applied Science: London, 1981.

(31) Kelly, B. T.; Marsden, B. J.; Hall, K. Irradiation Damage in Graphite due to Fast Neutrons in Fission and Fusion Systems.

(32) Telling, R. H.; Ewels, C. P.; El-Barbary, A. A.; Heggie, M. I. *Nat. Mater.* **2003,** *2,* 333-337.

(33) Kresse, G.; Furthmuller, J. *Comput. Mater. Sci.* **1996,** *6,* 15-50.

(34) Kresse, G.; Joubert, D. *Phys. Rev. B* **1999,** *59,* 1758-1775.

(35) Blochl, P. E.; Jepsen, O.; Andersen, O. K. *Phys. Rev. B* **1994,** *49,* 16223-16233.

(36) Henkelman, G.; Jonsson, H. *J. Chem. Phys.* **2000,** *113,* 9978-9985.

(37) Yao, F.; Gunes, F.; Ta, H. Q.; Lee, S. M.; Chae, S. J.; Sheem, K. Y.; Cojocaru, C. S.; Xie, S. S.; Lee, Y. H. *J. Am. Chem. Soc.* **2012,** *134,* 8646-8654.

(38) Fan, X. F.; Zheng, W. T.; Kuo, J. L.; Singh, D. J. *ACS Appl. Mat. Interfaces* **2013,** *5,* 7793-7797.

(39) Note, *the defect-induced band gap however will disappear in the dilute limit and the presence of defect will show up as a doping level instead.*

(40) Garay-Tapia, A. M.; Romero, A. H.; Barone, V. *J. Chem. Theory Comput.* **2012,** *8,* 1064-1071.

(41) Fan, X. F.; Zheng, W. T.; Kuo, J. L. *ACS Appl. Mat. Interfaces* **2012,** *4,* 2432-2438.

(42) Note, *thus for an AB stacking bi-layer graphene, the M-sites would be twice as many as the T-sites.*

(43) Sanville, E.; Kenny, S. D.; Smith, R.; Henkelman, G. *J. Comput. Chem.* **2007,** *28,* 899-908.

(44) Henkelman, G.; Arnaldsson, A.; Jonsson, H. *Comput. Mater. Sci.* **2006,** *36,* 354-360.

(45) Tang, W.; Sanville, E.; Henkelman, G. *J. Phys.: Condens. Matter* **2009,** *21,* 084204.

(46) Valencia, F.; Romero, A. H.; Ancilotto, F.; Silvestrelli, P. L. *J. Phys. Chem. B* **2006,** *110,* 14832-14841.

(47) Chan, K. T.; Neaton, J. B.; Cohen, M. L. *Phys. Rev. B* **2008,** *77,* 235430.

(48) Note, *For Li adsorption on pristine graphene or bilayer graphene, the upshift in the Fermi energy is ~1eV.*

(49) Zhou, L. J.; Hou, Z. F.; Wu, L. M. *J. Phys. Chem. C* **2012,** *116,* 21780-21787.

(50) Kubota, Y.; Ozawa, N.; Nakanishi, H.; Kasai, H. *J. Phys. Soc. Jpn.* **2010,** *79,* 014601.

(51) Chan, K. T.; Neaton, J. B.; Cohen, M. L. *Phys. Rev. B* **2008,** *77,* 235430.

(52) Note, *Those image sites are equivalently M-sites w.r.t. the top graphene sheet.*

(53) Toyoura, K.; Koyama, Y.; Kuwabara, A.; Oba, F.; Tanaka, I. *Phys. Rev. B* **2008,** *78,* 214303.

(54) Tran, T.; Kinoshita, K. *J. Electroanal. Chem.* **1995,** *386,* 221-224.

(55) Persson, K.; Sethuraman, V. A.; Hardwick, L. J.; Hinuma, Y.; Meng, Y. S.; van der Ven, A.; Srinivasan, V.; Kostecki, R.; Ceder, G. *J. Phys. Chem. Lett.* **2010,** *1,* 1176-1180.

**Graphic for table of contents (TOC figure):**

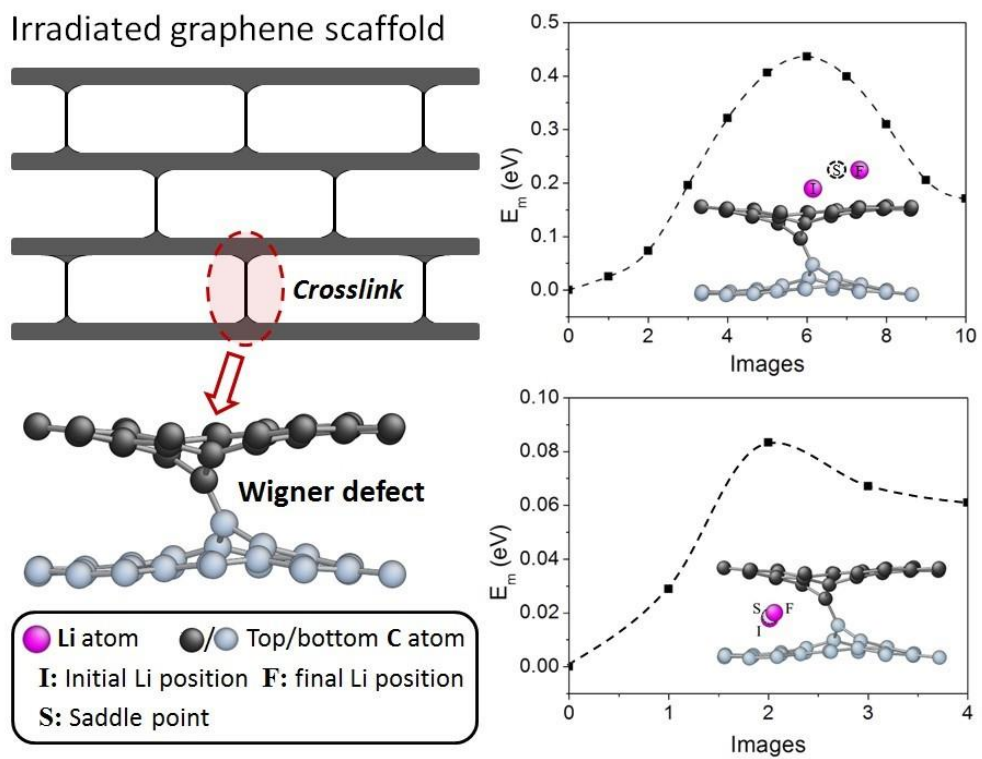